\documentclass{article}
\usepackage{arxiv}

\usepackage[utf8]{inputenc} 
\usepackage[T1]{fontenc}    
\usepackage{hyperref}       
\usepackage{url}            
\usepackage{booktabs}       
\usepackage{nicefrac}       
\usepackage{microtype}      
\usepackage{lipsum}		
\usepackage{doi}
\usepackage{graphicx}
\usepackage{amsmath}
\usepackage{amssymb,amsfonts}

\usepackage{cite}

\newcommand{\vvss}{``}
\newcommand{\vvdd}{''}
\newcommand{\vv}[1]{\vvss #1\vvdd}

\title{Implementation of an ITER-relevant QP-based Current Limit Avoidance algorithm in the TCV tokamak}

\author{
{Domenico Frattolillo}\\
DIETI, \\
Universit\`a degli Studi di Napoli Federico II, \\
and Consorzio CREATE, \\
Via Claudio 21, 80125 Napoli, Italy. \\
\texttt{domenico.frattolillo@unina.it} \\
\And
{Adriano Mele} \\
Swiss Plasma Center, \\ 
École Polytechnique Fédérale de Lausanne, \\
Rte Cantonale, 1015 Lausanne, Switzerland \\
\texttt{adriano.mele@epfl.ch} \\
\AND
{Cristian Galperti} \\
Swiss Plasma Center, \\
École Polytechnique Fédérale de Lausanne, \\
Rte Cantonale, 1015 Lausanne, Switzerland \\
\texttt{cristian.galperti@epfl.ch} \\
\And
{Luigi~E.~di~Grazia} \\
Consorzio CREATE, \\
Via Claudio 21, 80125 Napoli, Italy. \\
\texttt{luigiemanuel.digrazia@consorziocreate.it} \\
\AND
{Massimiliano Mattei} \\
DIETI, \\
Universit\`a degli Studi di Napoli Federico II, \\
and Consorzio CREATE, \\
Via Claudio 21, 80125 Napoli, Italy. \\
\texttt{massimiliano.mattei@unina.it} \\
\And
{Stefano Coda} \\
Swiss Plasma Center, \\
École Polytechnique Fédérale de Lausanne, \\
Rte Cantonale, 1015 Lausanne, Switzerland \\
\texttt{stefano.coda@epfl.ch} \\
\AND
{Gianmaria~De~Tommasi} \\
DIETI, \\
Universit\`a degli Studi di Napoli Federico II, \\
and Consorzio CREATE, \\
Via Claudio 21, 80125 Napoli, Italy. \\
\texttt{detommas@unina.it} \\
\And
{Alfredo Pironti} \\
DIETI, \\
Universit\`a degli Studi di Napoli Federico II, \\
and Consorzio CREATE, \\
Via Claudio 21, 80125 Napoli, Italy. \\
\texttt{pironti@unina.it} \\
\And
{Alessandro Tenaglia } \\
DICII, \\
Università degli Studi di Roma ``Tor Vergata'', \\
Via del Politecnico, 1, 00133 Roma, Italy. \\
\texttt{alessandro.tenaglia@uniroma2.it} \\
\AND
{Peter~deVries} \\
ITER Organization, \\
Route de Vinon sur Verdon, 13067 \\
St Paul Lez Durance, France \\
\texttt{Peter.DeVries@iter.org} \\
\And
{Luigi Pangione} \\
ITER Organization, \\
Route de Vinon sur Verdon, 13067 \\
St Paul Lez Durance, France \\
\texttt{Luigi.Pangione@iter.org} \\
\And
{Luca Zabeo} \\
ITER Organization, \\
Route de Vinon sur Verdon, 13067 \\
St Paul Lez Durance, France \\
\texttt{Luca.Zabeo@iter.org} \\
\AND
{TCV team} \\
See author list of B. P. Duval et al. \\
2024 Nucl. Fusion 64 112023 \\
\And
{Eurofusion Tokamak Exploitation Team} \\
See author list of E. Joffrin et al. \\
2024 Nucl. Fusion 64 112019 \\
}

\renewcommand{\shorttitle}{An ITER-relevant QP-based Current Limit Avoidance algorithm for the TCV tokamak}

\hypersetup{
pdftitle={Implementation of an ITER-relevant QP-based Current Limit Avoidance algorithm in the TCV tokamak},
pdfsubject={cs.SY, nucl-ex},
pdfauthor={D.~Frattolillo, A.~Mele, et al.},
pdfkeywords={Tokamak, Quadratic Programming, Optimization, Current Limit Avoidance},
}

\begin{document}
\maketitle
\begin{abstract}
    The problem of avoiding saturation of the coil currents is critical in large tokamaks with superconducting coils like ITER. Indeed, if the current limits are reached, a loss of control of the plasma may lead to a major disruption. Therefore, a Current Limit Avoidance~(CLA) system is essential to operate safely. This paper provides the first experimental evidence that the online solution of a constrained quadratic optimization problem can offer a valid methodology to implement a CLA. Experiments are carried out on the Tokamak à Configuration Variable (TCV) at the Swiss Plasma Center, showing the effectiveness of the proposed approach and its suitability for real-time application in view of future reactors such as ITER.
\end{abstract}


\section{Introduction}\label{section:Introduction}

\begin{figure}
    \centering
    \includegraphics[width = 0.75\linewidth]{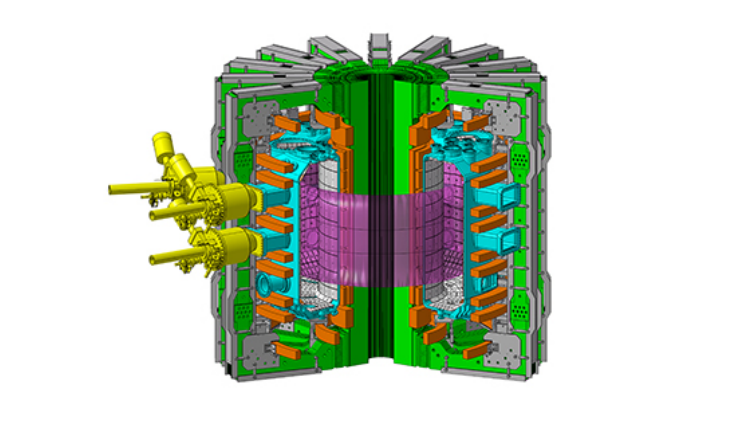}
    \caption{Cutaway view of the TCV tokamak, showing the PF coils (orange), the TF coils (green), the vacuum vessel (cyan), and the plasma region (purple). In addition, some of the ECRH launchers installed on TCV are also displayed in yellow.}
    \label{fig:tokamak}
\end{figure}

The tokamak concept is still considered a baseline for the design of a potential power plant reactor based on thermonuclear fusion~\cite{wesson2011tokamaks}.
The next step in this direction is ITER~\cite{itersite}, which will extensively test concepts and technologies required for the design and construction of a fusion reactor.
A tokamak is a toroidal-shaped device in which a fully ionized gas of hydrogen isotopes is heated up to temperatures of hundreds of millions of degrees Celsius, where the thermal agitation of the electrically charged nuclei can overcome the Coulomb repulsive force. This { enables} collisions producing fusion reactions. External magnetic fields allow the confinement of such a hot plasma. These fields are generated using currents flowing in the Poloidal Field~(PF) and the Toroidal Field~(TF) coils. These are shown in a simplified schematic for the TCV tokamak in Fig.~\ref{fig:tokamak}. An important role in tokamaks is played by magnetic control which is needed since day one and is essential to successfully control high-performance plasmas, such as those envisaged for~ITER~\cite{snipes2021iter}. 

Controlling the current induced in the plasma, as well as the plasma shape and position~\cite{AriolaPironti:Springer} are all problems relying on the currents flowing in the~PF circuits as actuators~\cite{de2019plasma}.
Robust control design techniques have been proposed in the literature~\cite{ariola2002design,mattei2013constrained,gerkvsivc2018model,wai2020control} aiming to reject unexpected disturbances and cope with model uncertainties. 
In practice however, any plasma magnetic control algorithm requires that PF coil currents are sufficiently far from their limits, in order to allow for reasonable margins to react to unexpected events. 
Open-loop strategies for the optimization of the nominal~PF current waveforms can be adopted~\cite{DIGRAZIA2022113027, mattei2006open,luo2020fast,di2024automated}. However, these may not be effective at the highest values of the plasma current or in the presence of unexpected disturbances~\cite{de2014shape}. 

As a consequence, it is important for a reliable plasma magnetic control system to include a Current Limit Avoidance (CLA) system to safely operate the machine when disturbances and/or uncertainties drive the~PF currents close to their limits~\cite{ambrosino2001control,varano2011performance,boncagni2011plasma}. 
{ In fact, for a critical infrastructure such as a nuclear power plant, common-mode failures, i.e. failures that occur simultaneously in several components or subsystems due to a shared cause, must be avoided. This can be done by adopting different alternative hardware and software solutions to improve the overall level of reliability of the system. To this end, in \cite{JOTA_CLA}, three different limit avoidance algorithms are described and tested through numerical simulations for the DEMO tokamak.
Two of these algorithms solve an online optimization problem, based on Quadratic Programming (QP) or Linear Programming (LP) respectively, while the third exploits a dynamic optimizer similar to what is proposed in~\cite{varano2011performance,CODIT:allocator}. These systems allow to modify the PF reference currents to satisfy the saturation limits while reducing the impact on the plasma shape by exploiting actuator redundancy.  
In this article, we aim at demonstrating the experimental feasibility of one of such solutions. In particular, a variant of the QP solution proposed in~\cite{JOTA_CLA} is implemented and experimentally validated on TCV.}

{Another possible approach which is often considered to limit currents in the active circuits is} Model Predictive Control~(MPC,~\cite{camacho2013model,mattei2013constrained}).
However, while the implementation of a~MPC scheme typically needs the knowledge of the plant dynamics, this is not the case for the CLA approach, provided that a so-called current control scheme is adopted~\cite{de2019plasma}. In addition, optimization problems arising in~MPC control schemes are usually larger than those solved in the~CLA formulation proposed in this article, as they take into account more than one prediction step. This makes the~CLA architecture less demanding from a computational point of view.

The CLA strategy has proven its effectiveness in numerical simulations on the~DEMO device~\cite{JOTA_CLA,donne2019european}, and has been proposed for inclusion in the ITER Plasma Control System (PCS)~\cite{DEVRIES2024114464}.
In this paper, the current limit avoidance system proposed for ITER is tested as a proof of concept on TCV, a medium-sized tokamak operated by the Swiss Plasma Center of the École Polytechnique Fédérale de Lausanne. 
The flexibility of TCV's upgraded Distributed Control System (Système de Contrôle Distribué, or SCD~\cite{galperti2024overview}) enables rapid and effective implementation and testing of novel plasma control solutions, including a recently implemented shape control algorithm~\cite{mele2024codit,mele:ssrn} and an alternative version of the CLA system which relies on a combination of a dynamic optimizer and a static annihilator~\cite{CODIT:allocator}. 
The significantly faster dynamics observed in TCV plasmas with respect to the ones expected in larger devices such as ITER require higher sampling frequencies in the control system, making the real-time implementation {of complex control algorithms} more challenging{. This} in turn implies that solutions adopted on TCV are expected to be easily scalable to future reactors, at least from the computational point of view, making TCV an ideal testbed for tokamak control technologies.
One of the main differences between ITER and TCV is in the technology of the coils. TCV is not a superconductive tokamak, which means that resistive copper coils are used to confine the plasma, while for~ITER a set of superconductive coils is foreseen. However, the proposed~CLA system only uses steady-state information, allowing for a proper validation of the proposed methodology despite the difference in the coil dynamics between the TCV and ITER devices. {Moreover, in TCV not all of the PF coil currents are controlled explicitly, but instead some combinations of currents are dedicated to plasma position and current control. This requires some adjustments in the QP-CLA strategy proposed in~\cite{JOTA_CLA}, which are discussed in section~\ref{section:CLA}.}

The rest of this paper is organized as follows. Section~\ref{section:Architecture} briefly describes the~TCV tokamak and its PCS; section~\ref{section:CLA} introduces the CLA methodology; in section~\ref{section:Exp}, experimental results are given; finally, section~\ref{section:Conclusions} concludes the article.

\section{TCV magnetic control system}\label{section:Architecture}

\begin{figure}
    \centering
    \includegraphics[width=0.5\linewidth]{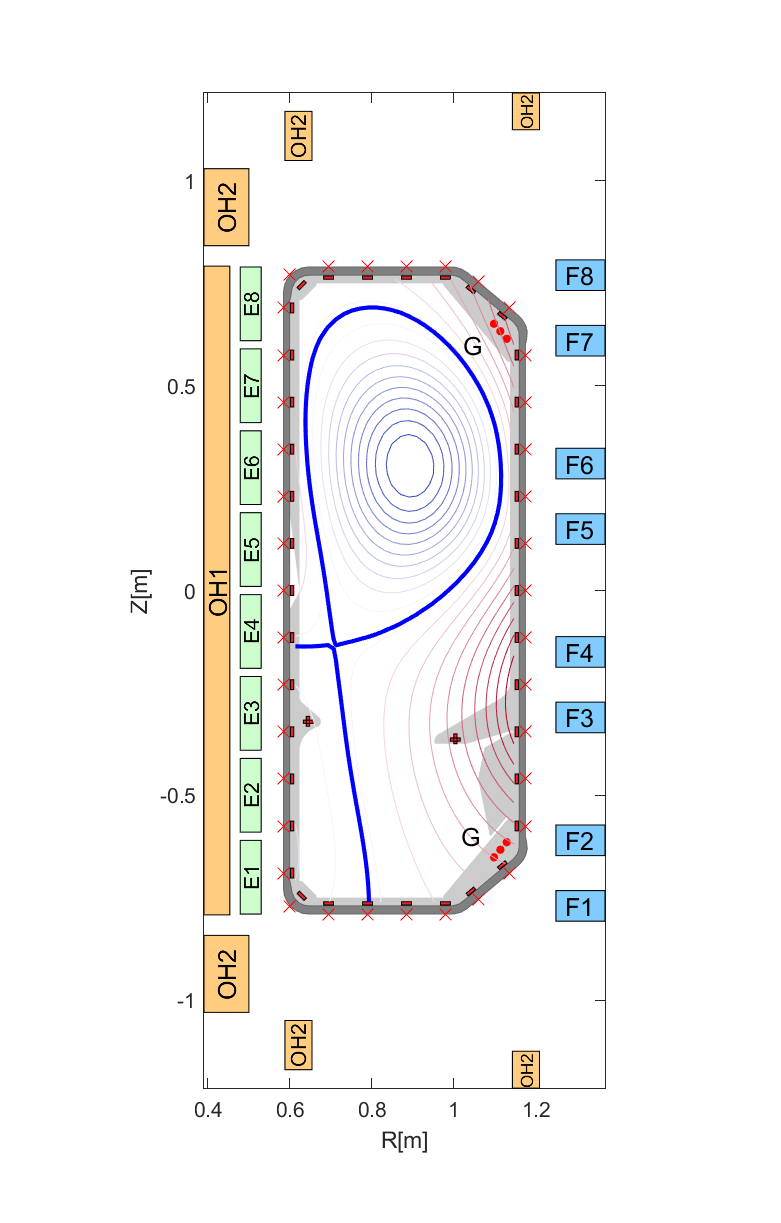}
    \caption{TCV poloidal cross-section (taken from~\cite{mele2024codit}).}
    \label{fig:tcv-layout}
\end{figure}

A layout of the TCV poloidal cross-section is shown in Fig.~\ref{fig:tcv-layout}.
The core magnetic controller in the TCV digital control system is designed to emulate the legacy analog system, called the \emph{hybrid} controller. The name comes from the fact that the original system was a hybrid digital-analog controller, where the computations were carried out analogically, but the gain matrices could be switched digitally.
The hybrid controller, represented in Fig.~\ref{figure:scheme}, is a Multiple-Input Multiple-Output (MIMO) controller that takes care of the vertical stabilization of the plasma and of the regulation of the plasma current $I_p$, of the radial and vertical plasma centroid displacement, and of tracking scenario coil currents. In particular, the average current in the OH coils and two combinations of the E and F ones are dedicated to $I_p$ control and to plasma position control, respectively. 
The remaining currents are projected in a space that is orthogonal to these combinations; {in particular, the difference $I_{DOH}=I_{OH,1}-I_{OH,2}$ is usually controlled to zero to minimize the stray field, while $14$ independent combinations of E and F currents, denoted by $I_{orth}$, are controlled explicitly (for more details, see~\cite{mele:ssrn})}. The hybrid controller generates voltage requests for the PF coils, which are summed to feedforward traces $V_{ff}$, computed offline, to obtain the voltage requests $V_a$ to the coil power supplies.

As is common in many tokamaks, the controller relies on magnetic and current measurements to reconstruct plasma position and current. In addition, a fast observer of the vertical position is obtained by using the in-vessel magnetic probes only, which are less affected by the shielding effect of the passive structures; this observer is used by the vertical stabilization control loop. 
Notice that, to be able to stabilize the plasma vertically, the digital version of this controller needs to run at a rather high sampling rate, typically chosen equal to $10$~kHz. 
The hybrid controller also includes a channel to control the plasma density, reconstructed through interferometry, which in practice is completely decoupled from the magnetic controller. 
An emulator of this controller is available in Matlab-Simulink, allowing to simulate the behavior of the controller before its real-time application. Moreover, the framework allows for automatic code generation and deployment of control algorithms directly in the SCD system~\cite{felici2014simulink}, providing a user-friendly environment to extend the capabilities of the original controller. 

Furthermore, the TCV magnetic control system is equipped with a plasma shape controller, as shown in Fig.~\ref{figure:scheme}; the design and validation of such controller are extensively discussed in~\cite{mele2024codit,mele:ssrn}. {The considered shape controller is based on the \emph{isoflux} approach~\cite{ferron1998}. To achieve the desired shape, a set of flux differences between control points specified by the user and a chosen boundary-defining point, such as an X-point or a limiter point, are controlled to zero. Moreover, the field components at the desired X-points are also controlled to zero in the case of diverted plasma configurations, while for limiter plasmas the component of the magnetic field tangent to the wall at the limiter point can be controlled to zero.}
The feedback signals for the shape controller, which modifies the reference signals of the hybrid block, are obtained by interpolating the poloidal flux and magnetic field maps obtained from the real-time equilibrium reconstruction code LIUQE~\cite{moret2015tokamak}. 
In practice, the necessity of a full equilibrium reconstruction limits the sampling rate of the shape controller to $1$~kHz.

Finally, note that the coils used for plasma shaping in TCV are the sets E1-8 and F1-8 shown in Fig.~\ref{fig:tcv-layout} in green and blue, respectively. Therefore, the associated currents $I_{EF}$ are the ones that will be considered in the allocation problem solved by the CLA.

\section{The Current Limit Avoidance Algorithm}\label{section:CLA}


\begin{figure}
\begin{center}
\includegraphics[width=\linewidth, trim=20 0 20 0]{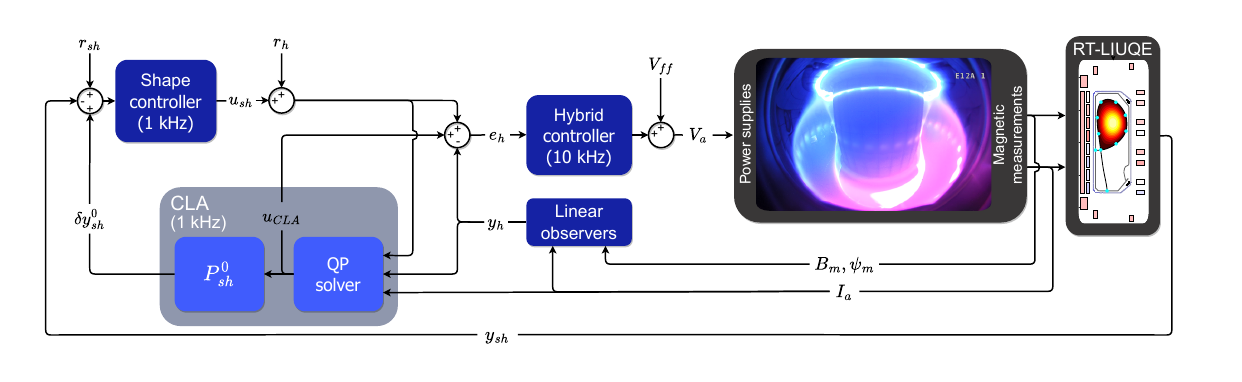}
\caption{{TCV control scheme, including the hybrid magnetic controller, the shape controller and the proposed CLA system (in light blue).}}\label{figure:scheme}
\end{center}
\end{figure}

The proposed CLA system, shown in {light} blue in Fig.~\ref{figure:scheme}, acts in parallel to the shape controller, providing additional control actions, summed to the hybrid controller reference signals $r_h$ with the aim of optimizing a quadratic cost function while simultaneously satisfying a set of hard linear constraints. Moreover, the steady-state effect of the allocator on the plasma shape is subtracted from the reference signals $r_{sh}$, to avoid any conflict between the two systems. 
The objective of the proposed CLA system is to reduce any excess of currents with respect to saturation limits by resorting to an alternative control pattern that allows the re-allocation of the currents while minimizing the degradation in terms of the shape control performance.


The CLA receives as input the references $r_h$ to the hybrid controller, modified by the action of the shape controller $u_{sh}$.
As {mentioned} in section~\ref{section:Introduction}, the proposed allocator acts based on the {expected} steady-state effect of the shape control actions on the coil currents. For this reason, the first step is to obtain an estimate of {this relation}. {Since, as explained in section~\ref{section:Architecture}, in TCV some coil current directions are not explicitly regulated, but are rather directly allotted to the control of the plasma position and current,} to obtain such an estimate the proposed~CLA system takes into account the static gain of the closed-loop system consisting of the linearized TCV plasma response in the vicinity of the considered plasma configuration and a state-space model of the hybrid controller. An estimate of said static gain can be computed based on linearized models obtained through the linearization module of the \texttt{fge} code~\cite{carpanese:phd}, included in the MEQ suite, similarly to what is done in \cite{mele2024codit,mele:ssrn,CODIT:allocator}.
{In this way, variations in the references to the plasma position channels can be translated into expected variations in the coil currents. It is worth to notice that, for the purposes of this article, the control channels associated to $I_p$ and to the current in the OH coils are considered as decoupled from those taking care of the plasma position and of the E and F coils, and as such they are not considered in the proposed architecture.}

Let us denote the static gain from (variations of) the hybrid control errors $e_h$ to the currents in the E and F coils of TCV as $P^0_{EF}$ and the gain from $e_h$ to variations in the plasma shape descriptors $y_{sh}$ as $P^0_{sh}$. 
{ 
The vector $e_h$ includes control errors on the $I_{orth}$ combinations of the E and F coil currents introduced in section \ref{section:Architecture} and on the plasma centroid position. The shape descriptors $y_{sh}$, instead, include the magnetic field components at the desired X-points and the differences between poloidal flux at a set of user-specified contour points and a boundary-defining point (e.g. the X-point for the diverted configuration considered in the examples), according to the well-established isoflux control paradigm. 
}
%

Assuming that the plasma is close to the desired scenario configuration and denoting with $y_h$ the outputs of the plant controlled by the hybrid loop, an estimate of the steady-state value of the currents in the E and F coils of TCV can be obtained at each controller step as
\begin{equation}\label{eq:Iss}
I^0_{EF} = \underbrace{I_{EF}}_{\substack{\text{measured} \\ \text{currents}}} + \,\,
\underbrace{
    P^0_{EF} (u_{sh} + {u_{CLA}} + r_h - y_h) 
}_{\substack{\text{expected variation} 
\\ \text{at steady-state}}} 
\,,
\end{equation}
%
{For simplicity, with reference to Fig.~\ref{figure:scheme}, in what follows we will assume $V_{ff}=const$ during the shape control phase. This is not the case in standard TCV shots, but the voltage feedforwards were frozen to their value at the controller activation times for the experiments reported in this article.
Note that the introduction of the static gain matrix $P_{EF}^0$ is an approximation used to adapt the logic proposed in~\cite{JOTA_CLA} to TCV, where the control channels for the PF coil currents and for the plasma position are not independent.
To this end, the proposed adaptation relies on the assumption that the hybrid controller is accurate enough to bring the controlled errors close to zero at steady-state. In practice, this means that, after the transient phase, $I_{EF}^0 \approx \lim_{t\to\infty} I_{EF}$ in eq.~\eqref{eq:Iss}. Note that if the hybrid controller is capable of accurate reference tracking and if the plasma configuration is not significantly perturbed by the action of the allocator, then the term in round brackets in~\eqref{eq:Iss} is expected to be small and the linearized response approximation holds. 
}

{
It is worth remarking however that the proposed architecture in practice only acts on the references to the inner control loops, and hence it is only accurate to the extent to which the underlying control loop (the hybrid controller in this case) is. 
In fact, if a steady-state error was present in the quantities tracked by such a controller, this would be reflected in the steady-state currents achieved, i.e. $I_{EF}^0\neq \lim_{t\to\infty}I_{EF}$. More details regarding this point are discussed in section~\ref{subsection:TCII}.
}

{
On the other hand, it is also worth to note that, if an explicit control of the PF coil currents is used (as foreseen in ITER's magnetic control architecture), the proposed allocator can directly act on the references to such controller, and hence the approximate estimate of the steady-state currents given by~\eqref{eq:Iss} is not necessary. This is what is done, for example, in the approach proposed in~\cite{JOTA_CLA}, where the CLA acts on the current references fed to an inner control loop for the coil currents. As it happens with the solution proposed in the present work, the allocator in~\cite{JOTA_CLA} is also agnostic with respect to the steady-state error of the coil current controller.
In practice, the introduction of the static gain $P_{EF}^0$ in~\ref{eq:Iss} is a device used to adapt the logic proposed for ITER in~\cite{JOTA_CLA} to TCV, where the control channels for the PF coil currents and for the plasma position are not independent.
}
%

%

{We are interested in the variations of the shape descriptors and the EF currents induced by the action of the proposed allocation strategy once a steady-state has been reached. These are denoted as $\delta y^0_{sh}$ and $\delta I^0_{EF}$, respectively. The steady-state effect of the CLA on the plasma shape descriptors $y_{sh}$ and on the EF coil currents can be estimated as
\begin{equation}  \label{eq:QPss}
  \delta y^0_{sh} = P^0_{sh} u_{CLA} \,, \quad \delta I^0_{EF} = P^0_{EF} u_{CLA} \,.
\end{equation}
}
At every time step, the CLA solves the following Quadratic Programming (QP) optimization problem
%
{
\begin{equation}  \label{eq:CLA-opt}
    \begin{aligned}
        &\min_{u_{CLA}} \qquad &\delta y^{0T}_{sh} W \delta y^0_{sh} + \delta I^{0T}_{EF} Q \delta I^0_{EF} \\
        &\text{s.t.}  &\underline{I}_{EF} \le I^0_{EF} \le \overline{I}_{EF}
    \end{aligned}
\end{equation}
}
where $\underline{I}_{EF}$ and $\overline{I}_{EF}$ represent the lower and upper bounds on the $EF$ coil currents, respectively.
The first term of the cost function aims at minimizing the effect of the~CLA on the shape descriptors $y_{sh}$, while the second aims at reducing its effect on the PF coil currents. { The weights $W$ and $Q$ are properly scaled to consider the magnitude difference between shape and active currents values.} Note that, if the steady-state current vector $I^0_{EF}$ satisfies the constraints of the optimization problem, the optimal solution is given by~$u_{CLA}~=~0$,~i.e. the~CLA leaves the references to the hybrid controller~$r_h$ unmodified. 
\footnote{On the other hand, the operator may decide to minimize the $I_{EF}$ steady-state current norm by changing the second term in the optimization to $I^{0T}_{EF} Q I^0_{EF}$, similarly to what is proposed in~\cite{CODIT:allocator}. In this case, typically a relatively low weight $Q$ is assigned to the current term with respect to the weight $W$ assigned to the shape variations, and the optimizer will attempt to rearrange the currents in order to obtain a comparable shape accuracy while minimizing the weighted current norm as a secondary objective.} 

{
Substituting equations~\eqref{eq:Iss} and \eqref{eq:QPss} into \eqref{eq:CLA-opt} we obtain the following standard QP problem with linear constraints
\begin{equation}  
    \begin{aligned}
    &\min_{u_{CLA}} \qquad & (P^0_{sh} u_{CLA})^T W  (P^0_{sh} u_{CLA}) + (P^0_{EF} u_{CLA})^T Q (P^0_{EF} u_{CLA}) \\
    &\text{s.t.}  \qquad 
        & \begin{aligned}
            +P^0_{EF}u_{CLA} &\le +\bar{I}_{EF}       - \left(I_{EF} + P^0_{EF} (u_{sh} + r_h - y_h) \right) \\
            -P^0_{EF}u_{CLA} &\le -\underline{I}_{EF} + \left(I_{EF} + P^0_{EF} (u_{sh} + r_h - y_h) \right)
        \end{aligned}
    \end{aligned}
\end{equation}
}

On the other hand, since the current saturations are specified as hard inequality constraints, if some of the currents in the vector $I^0_{EF}$ exceed the limit, then the~CLA will provide a nonzero correction $u_{CLA}$ to the requests $r_h$ given to the hybrid controller, to bring all currents back to the safe region while minimizing at the same time the effect on the plasma shape. 
In addition, as shown in Fig.~\ref{figure:scheme}, a term $P^0_{sh} u_{CLA}$ is subtracted from the reference signals $r_{sh}$, in order to hide the effect of the allocator at steady state from the point of view of the shape controller. In this way, the outer shape control loop will not react to the changes made by the~CLA. 

An important remark is that in the case of TCV and differently to what has been discussed in section~\ref{section:Introduction}, the design of the CLA still (loosely) depends on the knowledge of the structure of the hybrid controller, and in particular of its steady-state gain. This is due to the fact that, in TCV, the control of the coil currents and the plasma current and position is tightly coupled, while the shape control also generates reference variations for the plasma position channels. On the other hand, in other tokamak magnetic control architectures, where the plasma position and $I_p$ controllers generate references for an internal loop that regulates the currents in the PF coils, this knowledge is not necessary under the assumption that the coil current controller has an approximately unitary steady-state gain on all channels, i.e., that the steady-state values of the PF currents are close to the respective references. This is, for example, the case of the foreseen ITER control system. 
{As a consequence, the implementation of the algorithm is also expected to be simpler, as the effect of the position controller on the PF coil currents must not be taken explicitly into account. 

However, it is important to note that the performance of the proposed allocation strategy, which acts similarly to a reference governor and relies on the availability of a model of the relationship between the PF currents and the plasma shape, strongly depends on the accuracy of the available models and on the underlying control architecture.
As the codes used to produce the results in this paper are available to the fusion community, there is no reason to expect lower modeling accuracy on ITER compared to TCV. Further details on the impact of the accuracy of the underlying controller on the performance of the proposed strategy are given instead in section~\ref{subsection:TCII}.}

Finally, it should be mentioned that, in other formulations of the same problem (such as~\cite{JOTA_CLA}), the effect of the CLA on the plasma current is also taken into account explicitly through an additional term in the optimization cost function. However, in TCV the $I_p$ control is achieved by a dedicated set of coils (OH1-2 in Fig.~\ref{fig:tcv-layout}) on time scales that are significantly faster than those of the shape control, so this term can be neglected in the case under exam.

\section{Experimental results}\label{section:Exp}
The effectiveness of the proposed CLA approach is demonstrated in three different test cases on a TCV lower single-null reference scenario, shown in Fig.~\ref{fig:tcv-layout}.
{For all the considered cases, saturations of $\pm 3.5$~kA have been imposed on all E and F coils. Moreover, the following situations have been considered:}
\begin{itemize}
    \item \textbf{Test case I}: the limit on one circuit is fictitiously reduced {to $2.5$~kA}.
    \item \textbf{Test case II}: the limits on three circuits are fictitiously reduced {to $2.5$~kA}.
    \item \textbf{Test case III}: {like test case II, but} the limit on one of the circuits is fictitiously reduced in a time-varying fashion.
\end{itemize}
%

\begin{figure}
\begin{minipage}[c]{0.45\linewidth}
\includegraphics[width=\linewidth]{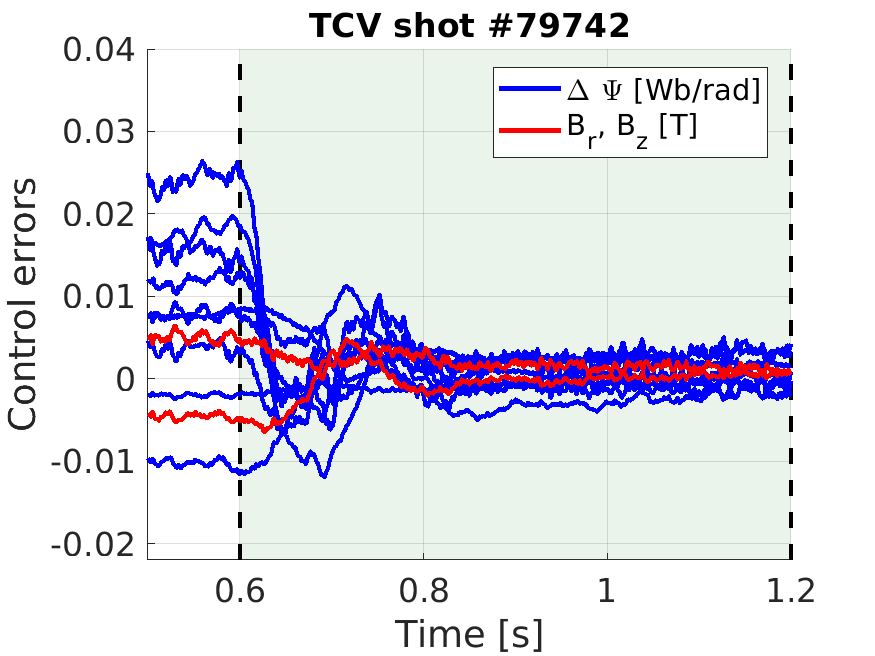}
\caption{Reference scenario {(TCV pulse \#79742)} - time traces of the isoflux control errors. {The poloidal flux errors (at contour and strike points) are shown in blue, while the magnetic field errors (at the X-point) are in red.} The beginning and end of the controlled time window are indicated by the dashed black lines.}\label{figure:shapeerrR}
\end{minipage}
\hfill
\begin{minipage}[c]{0.52\linewidth}
\includegraphics[width=\linewidth]{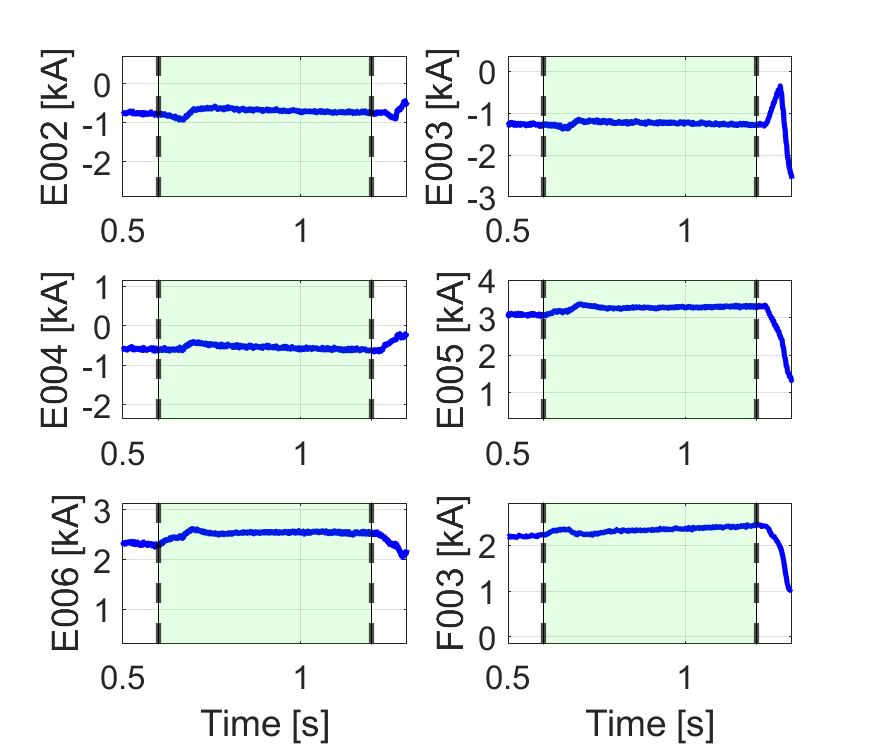}
\caption{Reference scenario - time traces for some of the currents in the E and F coils.
}\label{figure:currentR}
\end{minipage}%
\end{figure}

The reference scenario, taken from TCV pulse \#79742, is a $I_p = 250$~kA lower single null plasma. The isoflux shape control described in~\cite{mele:ssrn} is activated during the flat top phase from $0.6$~s to $1.2$~s. 
Fig.~\ref{figure:shapeerrR} shows the time behavior of the shape control errors in the case where no saturation limit is acting on the~PF currents. Fig.~\ref{figure:currentR} shows the currents in some circuits of interest for the considered test cases. 

{
With reference to the optimization problem~\eqref{eq:CLA-opt}, diagonal $W$ and $Q$ weighing matrices have been used. The individual weights in $W$ can be used to promote accuracy on specific channels, for instance on the control of radial gaps for configurations which are very close to the inner or outer walls. In the proposed examples, the same weights used in the shape controller were kept, i.e. a weight of $2$ on the strike points and of $1$ on all the other channels. The weight for each control point is then scaled proportionally to the inverse of the gradient of the associated controlled quantity, to \vv{equalize} the shape accuracy in terms of local boundary deformations.

For what concerns the weights in $Q$ instead, a unitary gain was simply chosen on all channels. A further scalar relative weight between $W$ and $Q$ can be introduced to empirically regulate the time scale on which the other currents are rearranged to compensate for the ones affected by the hard constraints. In practice, a lower weight on $Q$ relative to $W$ means that the allocator has more freedom to modify the requested currents in order to minimize the projected shape error. 
A first guess for such a scaling can be obtained by looking at the values of the typical shape control errors and of the reference currents in a similar discharge, but then it is usually tuned empirically based on offline simulations. For the proposed examples, we multiplied the original $W$ matrix by a factor $\times10^3$.
}

The real-time implementation of the CLA system relies on the availability of an effective solver for the QP problem~\eqref{eq:CLA-opt}. The implementation proposed for this work is based on the \texttt{C++} library~\cite{quadprogPP}, which implements the Goldfarb-Idnani active-set dual method~\cite{goldfarb1983numerically}. As it can be seen from figs.\ref{figure:itI}-\ref{figure:itII}-\ref{figure:itIII},  the solver converged within~3 iterations for all the considered cases, even when multiple current saturations were hit simultaneously.

\subsection{Test Case I}\label{subsection:TCI}
\begin{figure}
\begin{center}
\includegraphics[width = 0.8\linewidth]{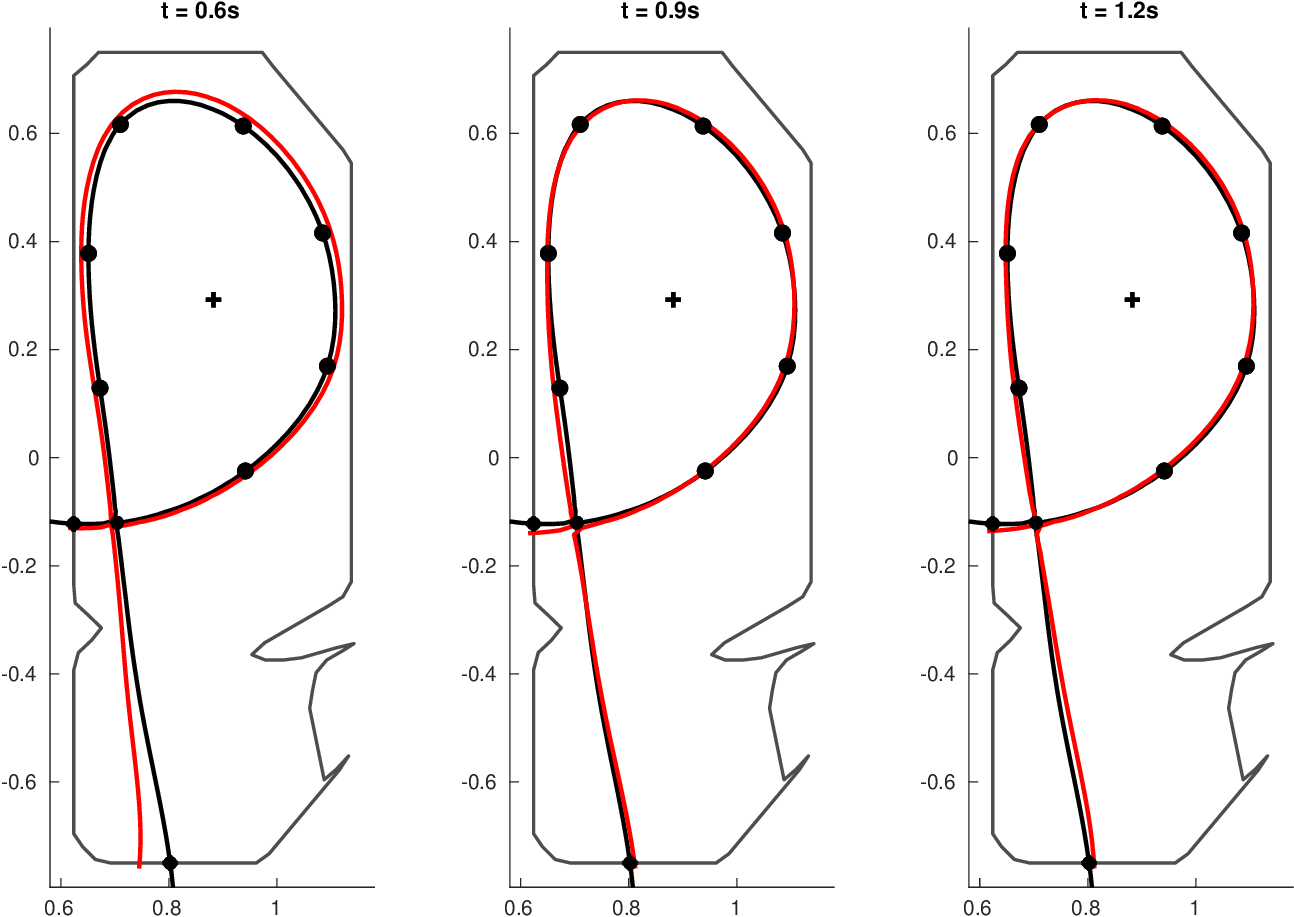}
\caption{Test Case I -  snapshots at different time instants. The experimental plasma separatrix is shown in red, while the boundary for the reference target equilibrium is shown in black. The black dots are the considered isoflux control points.}\label{figure:shapeI}
\end{center}
\end{figure}

\begin{figure}
\begin{minipage}[c]{0.45\linewidth}
\includegraphics[width=\linewidth]{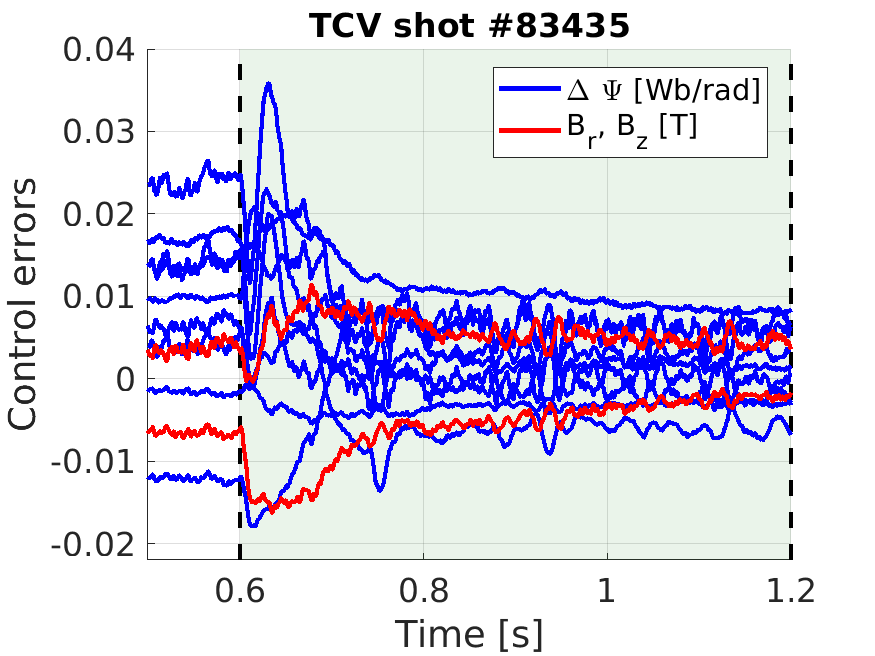}
\caption{Test Case I {(TCV pulse \#83435)} - time traces of the isoflux control errors. {The poloidal flux errors (at contour and strike points) are shown in blue, while the magnetic field errors (at the X-point) are in red.} The beginning and end of the controlled time window are indicated by the dashed black lines.}\label{figure:shapeerrI}
\end{minipage}
\hfill
\begin{minipage}[c]{0.52\linewidth}
\includegraphics[width=\linewidth]{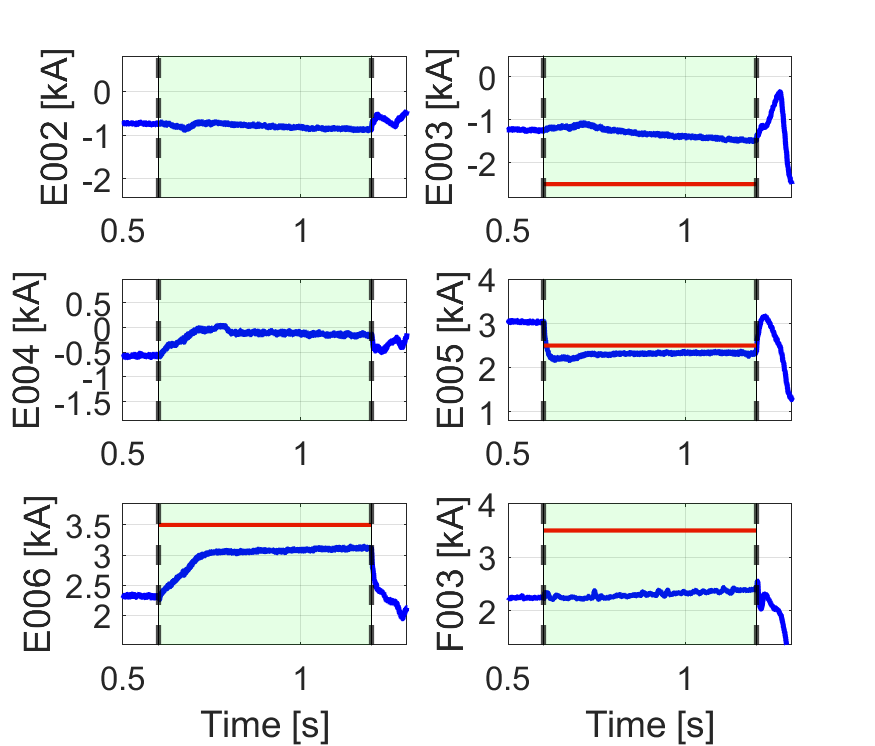}
\caption{Test Case I - time traces of the PF currents (blue) compared with saturation limits (red).}\label{figure:currentI}
\end{minipage}%
\end{figure}

\begin{figure}
\begin{minipage}[c]{0.48\linewidth}
\includegraphics[width=\linewidth]{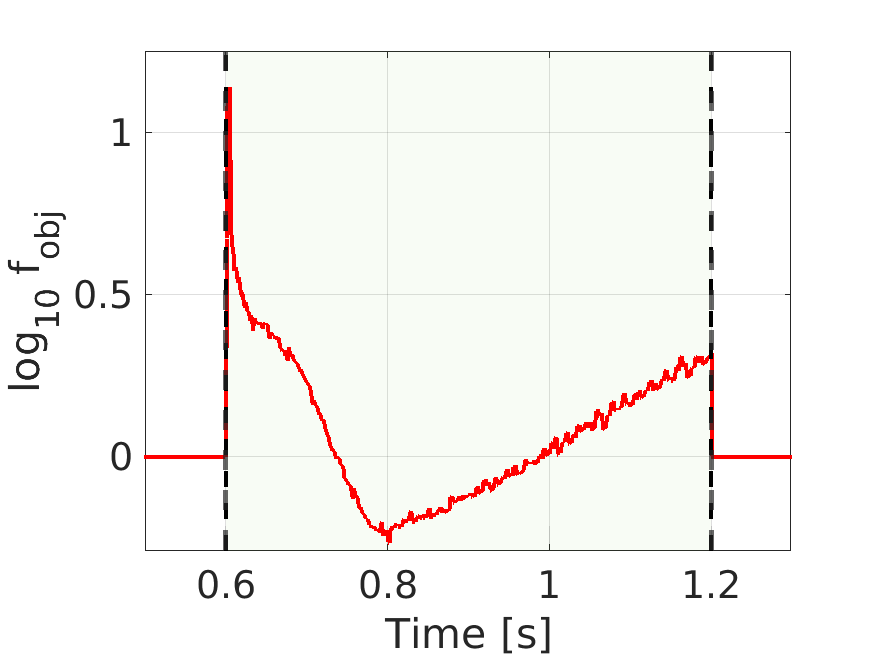}
\caption{Test Case I - time trace of the QP problem objective function.}\label{figure:fobjI}
\end{minipage}
\hfill
\begin{minipage}[c]{0.48\linewidth}
\includegraphics[width=\linewidth]{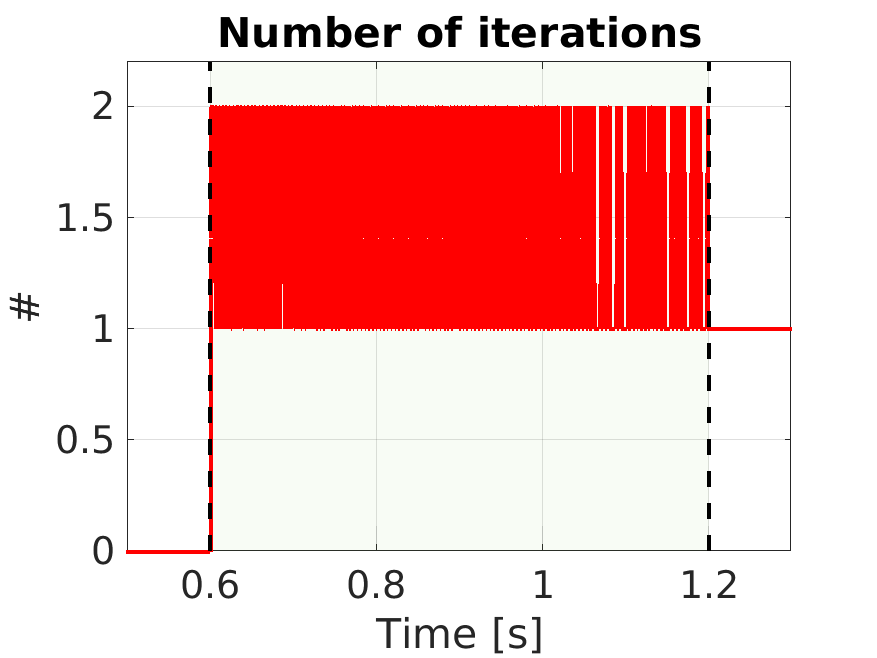}
\caption{Test Case I - iterations performed by the QP solver.}\label{figure:itI}
\end{minipage}%
\end{figure}

The first of the proposed tests was performed in TCV shot \#83435, and the results are reported in figs.~\ref{figure:shapeI}-\ref{figure:fobjI}. 
The shape control error traces are shown in Fig.~\ref{figure:shapeerrI}, where a degradation of the control performance can be observed with respect to the baseline case in Fig.~\ref{figure:shapeerrR}.
On the other hand, Fig.~\ref{figure:shapeI} shows the snapshots at three time instants of interest, from which it can be seen that the degradation in plasma shape control performance introduced by the CLA action is almost negligible at steady-state from the practical point of view. 
The coil currents that are closest to saturation and the corresponding saturation limits are also reported in Fig.~\ref{figure:currentI}, where it can be seen how the E5 current returns into an acceptable band following the activation of the CLA at $t=0.6s$. The QP cost function is reported in Fig.~\ref{figure:fobjI}, while the iterations needed by the QP solver are shown in Fig.~\ref{figure:itI}.

\subsection{Test Case II}\label{subsection:TCII}
\begin{figure}
\begin{center}
\includegraphics[width = 0.8\linewidth]{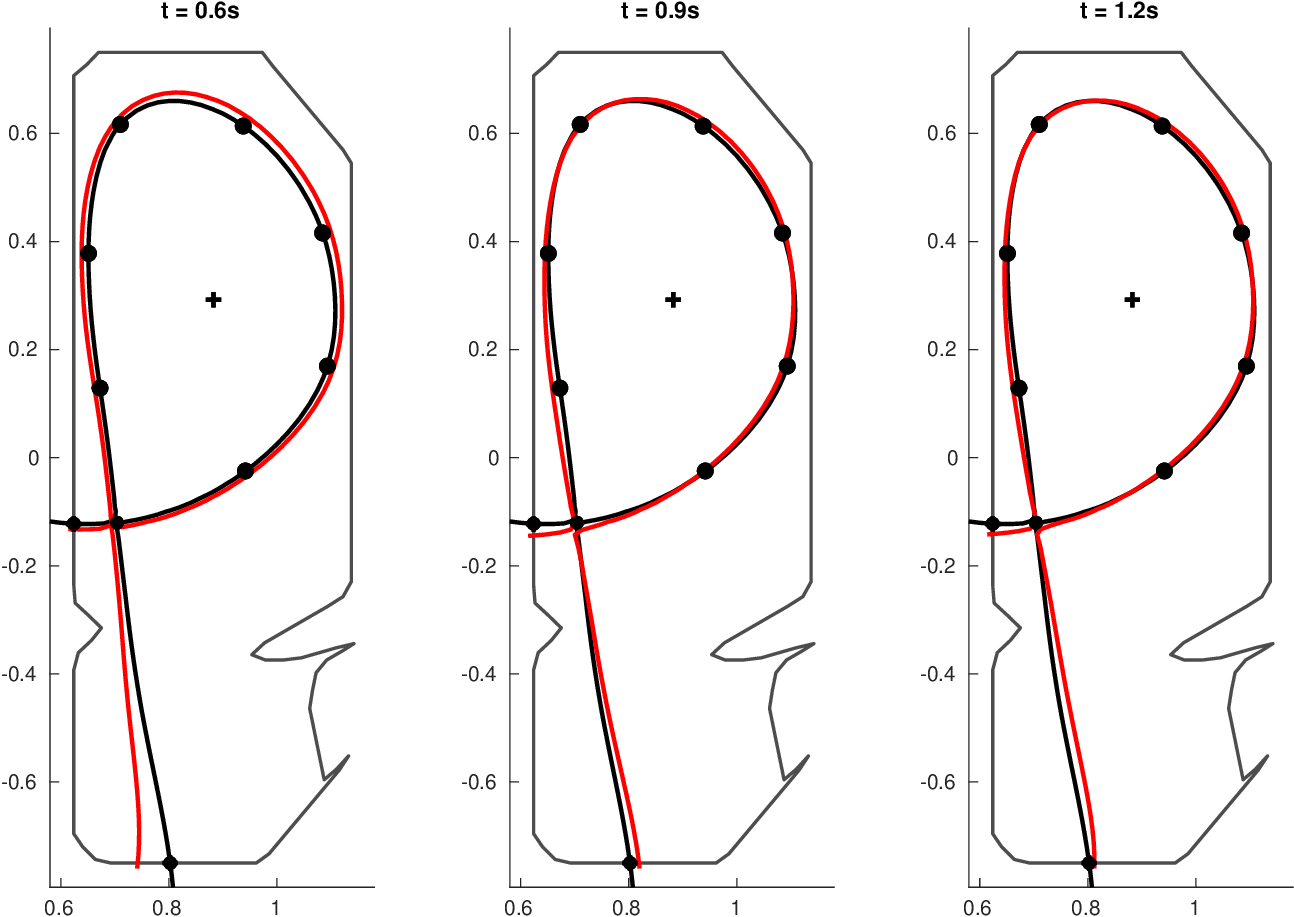}
\caption{Test Case II - snapshots at different time instants. The experimental plasma separatrix is shown in red, while the boundary for the reference target equilibrium is shown in black. The black dots are the considered isoflux control points.}\label{figure:shapeII}
\end{center}
\end{figure}

\begin{figure}
\begin{minipage}[c]{0.45\linewidth}
\includegraphics[width=\linewidth]{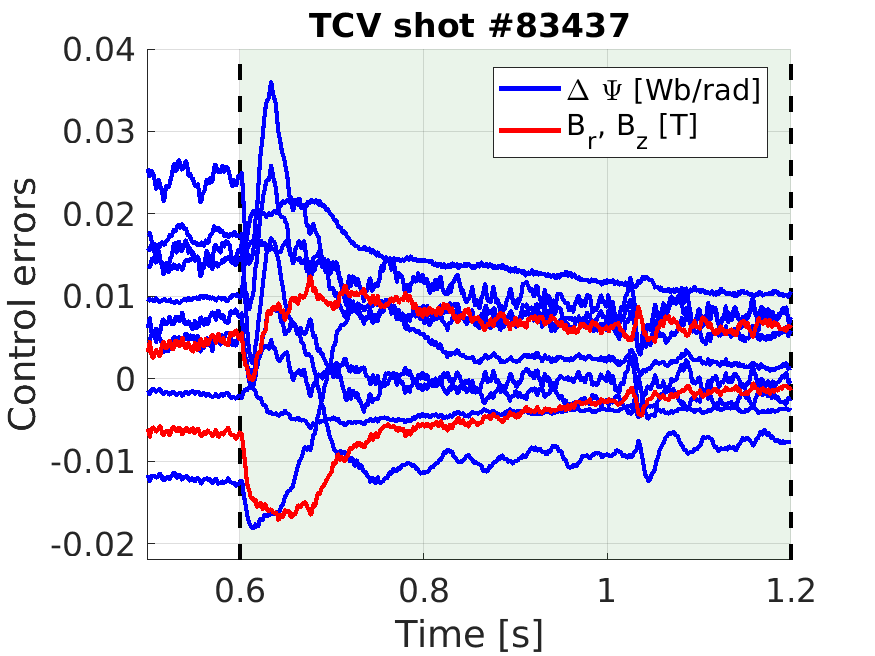}
\caption{Test Case II {(TCV pulse \#83437)} - time traces of the isoflux control errors. {The poloidal flux errors (at contour and strike points) are shown in blue, while the magnetic field errors (at the X-point) are in red.} The beginning and end of the controlled time window are indicated by the dashed black lines.}\label{figure:shapeerrII}
\end{minipage}
\hfill
\begin{minipage}[c]{0.52\linewidth}
\includegraphics[width=\linewidth]{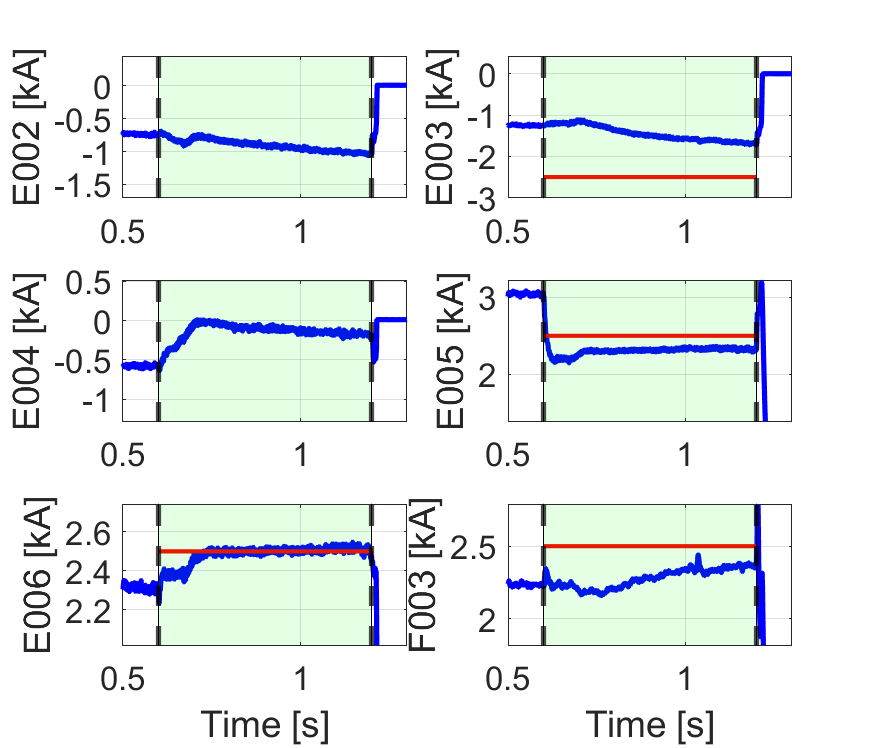}
\caption{Test Case II - time traces of the PF currents (blue) compared with saturation limits (red).}\label{figure:currentII}
\end{minipage}%
\end{figure}

\begin{figure}
\begin{minipage}[c]{0.48\linewidth}
\includegraphics[width=\linewidth]{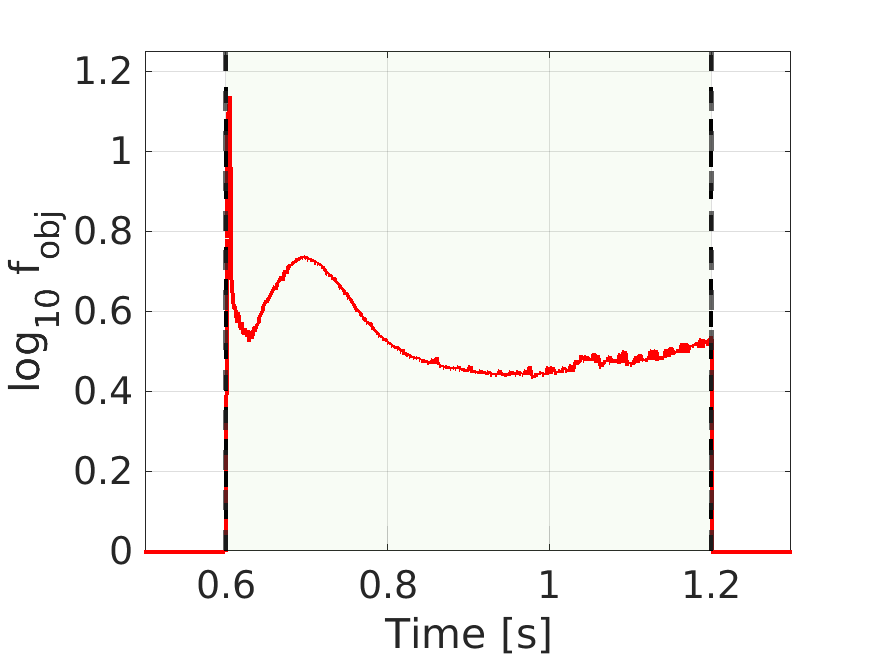}
\caption{Test Case II - time trace of the QP problem objective function.}\label{figure:fobjII}
\end{minipage}
\hfill
\begin{minipage}[c]{0.48\linewidth}
\includegraphics[width=\linewidth]{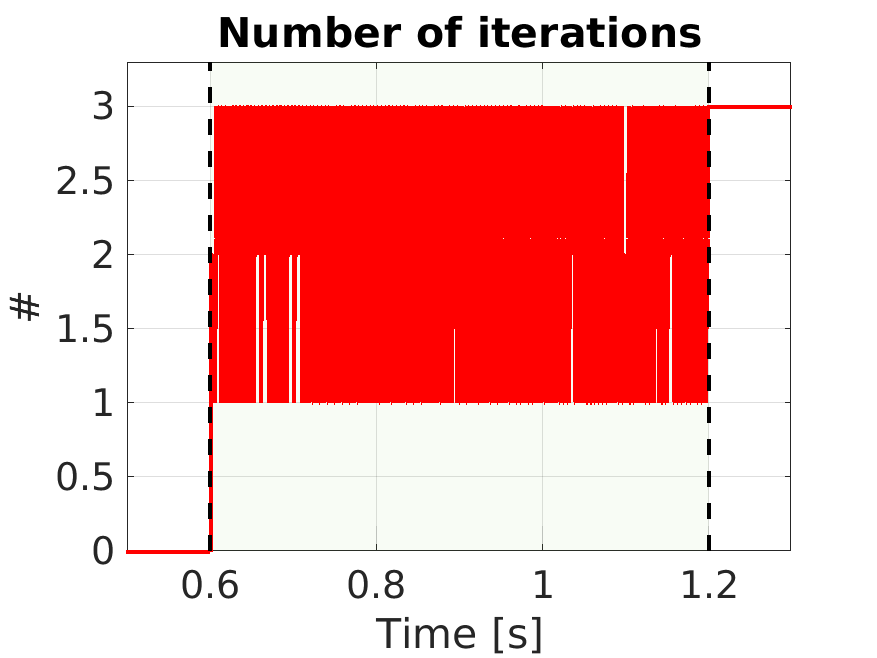}
\caption{Test Case II - iterations performed by the QP solver.}\label{figure:itII}
\end{minipage}%
\end{figure}

The results for the second test case, performed in TCV shot \#83437, are reported in Figs.~\ref{figure:shapeII}-\ref{figure:fobjII}. The snapshots at three different time instants in Fig.~\ref{figure:shapeII} illustrate the plasma shape degradation due to the CLA contribution; it can be observed how, in this case, the performance degradation is more evident in the region close to the X-point and the inner strike point. Fig.~\ref{figure:shapeerrII} displays the time traces of the shape control errors.
The relevant coil currents and corresponding saturation limits are reported in Fig.~\ref{figure:currentII}, where it can be seen how the E5 current is brought back into an acceptable band after the activation of the CLA at $t=0.6$~s; to compensate, the E6 current moves closer to its saturation, also artificially reduced to $2.5$~kA. The QP cost function is reported in Fig.~\ref{figure:fobjII}, with the iterations needed by the QP solver shown in Fig.~\ref{figure:itII}.

\begin{figure}[h]
    \centering
    \includegraphics[width=\linewidth]{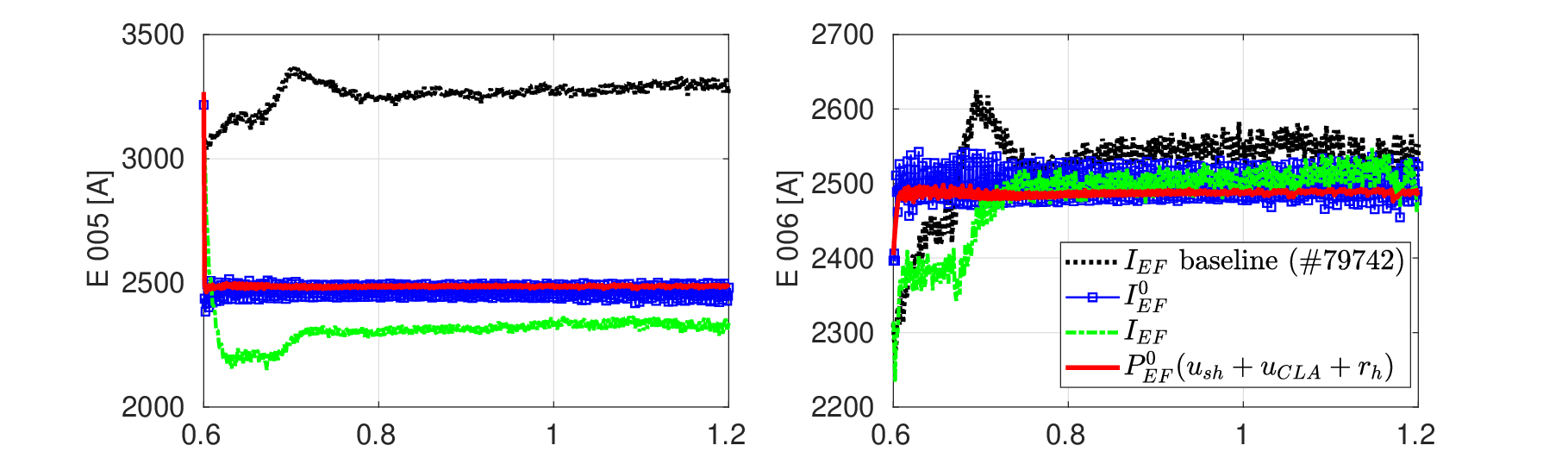}
    \caption{TCV pulse \#83437. In green the measured $I_{EF}$ currents, in blue the projected steady-state values $I_{EF}^0$, in red the reference to the hybrid controller modified by the action of the shape control and of the CLA system and projected through the steady-state gain $P_{EF}^0$. In black the same currents in the reference pulse \#79742.}
    \label{fig:IE56}
\end{figure}

\begin{figure}[h]
    \centering
    \includegraphics[width=\linewidth]{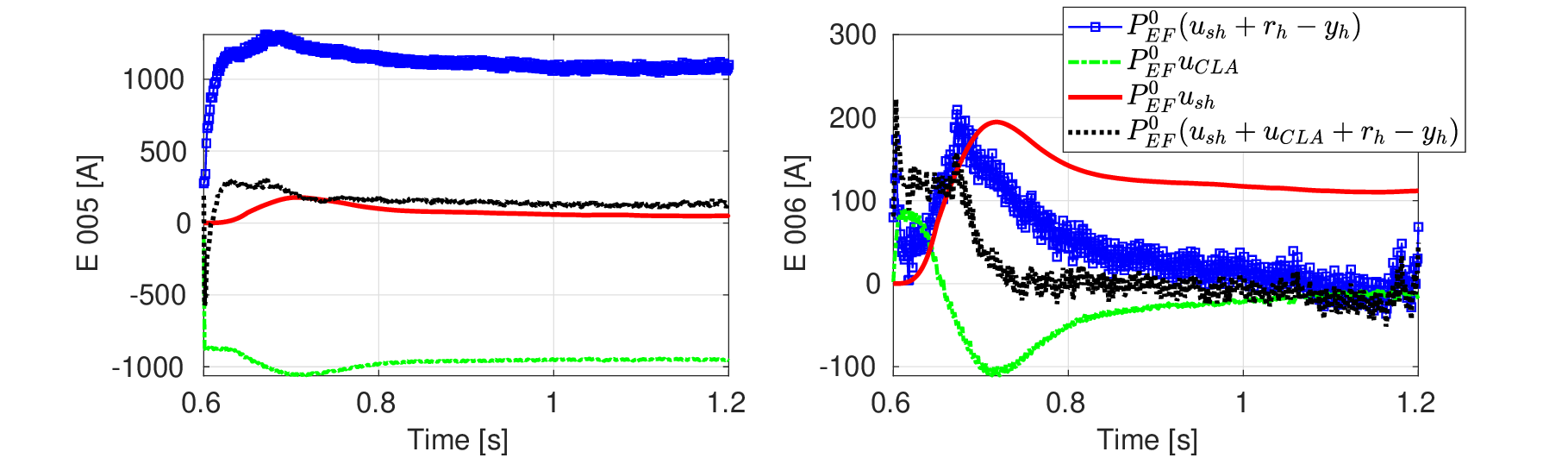}
    \caption{TCV pulse \#83437. In black the tracking error on the currents in the E005 (left) and E006 (right) coils. In green and red the reference modifications introduced by the shape controller and the allocator respectively.}
    \label{fig:ref56}
\end{figure}

{
We take advantage of this example to illustrate how the steady-state accuracy of the underlying control loop affects the precision of the imposed saturation limits, as discussed in section~\ref{section:CLA}. 
Consider the currents in coils E005 and E006. These coils are not used for plasma position control. Hence, their currents are directly controlled by hybrid and the respective entries in the static-gain matrix $P_{EF}^0$ are approximately unitary from reference to corresponding current and zero elsewhere, i.e. for these control channels $P^0_{EF} r_h \approx r_h$.

Fig.~\ref{fig:IE56} shows the currents in these coils for both the \vv{baseline} scenario (pulse \#79742, dashed black) and the considered test case (green). It can be seen how the current on E005 is significantly reduced with respect to the reference case, and successfully brought below the saturation threshold. The current in E006 is also reduced, but it is still slightly above the saturation limit (by approximately $50$~A, i.e. $\sim2\%$ of its value). However, the projected steady-state reference value $P_{EF}^0 (r_h + u_{sh} + u_{CLA})$ is correctly lower than the imposed saturation limit of $2500$~A. The same is true for the $I_{EF}^0$ quantity, which is approximately equal to the saturation limit (minus the effect of high frequency noise, coming from the measured $I_{EF}$ values in~\eqref{eq:Iss}. Similarly, for coil E005, the expected $I_{EF}^0$ value (blue trace) is exactly at the saturation limit, while the achieved current has a small offset, this time in a beneficial direction, of approximately $150$~A ($\sim6\%$). As discussed above, for both cases the effect of the projection matrix $P_{EF}^0$ is negligible. As a consequence, the residual error between the modified reference $r_h + u_{sh} + u_{CLA}$ (red trace) and the achieved current (green trace) should be attributed to the accuracy of the hybrid controller.

For completeness, in Fig.~\ref{fig:ref56} the effects of the different terms in~\eqref{eq:Iss} are considered. 
In particular, the panels show the (projected) E005 and E006 current errors with the allocator and shape controller actions (dotted black lines). Moreover, the individual contributions to the modified references due to the shape control and the allocator are shown in red and green, respectively. 
Again, it is worth to remark that the action of the effect of the projector $P_{EF}^0$ is negligible for these currents, hence these traces can be considered in practice as error or reference signals on the coil currents themselves. 
Looking at the case of E005 (left plot), it can be seen for instance how the allocator introduces an offset at steady-state of about $-950$~A, which brings the current back into the saturation limits. Moreover, the error between the modified reference and the hybrid feedback signal is reported in black. Its nonzero steady-state value is also reflected in the small offset between the green and red traces in Fig.~\ref{fig:IE56}
}

\subsection{Test Case III}\label{subsection:TCIII}
\begin{figure}
\begin{center}
\includegraphics[width = 0.8\linewidth]{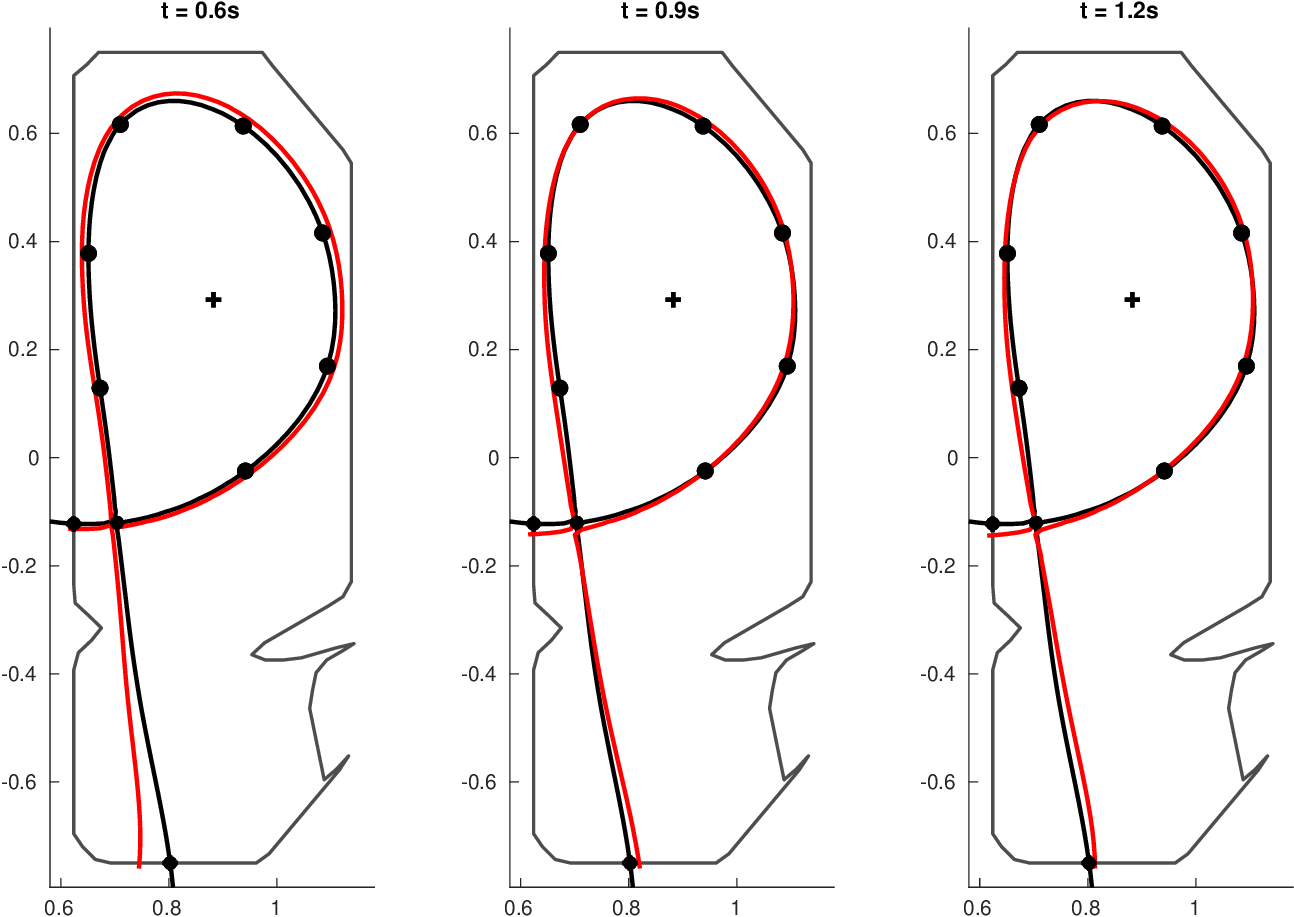}
\caption{Test Case III -  snapshots at different time instants. The experimental plasma separatrix is shown in red, while the boundary for the reference target equilibrium is shown in black. The black dots are the considered isoflux control points.}\label{figure:shapeIII}
\end{center}
\end{figure}

\begin{figure}
\begin{minipage}[c]{0.45\linewidth}
\includegraphics[width=\linewidth]{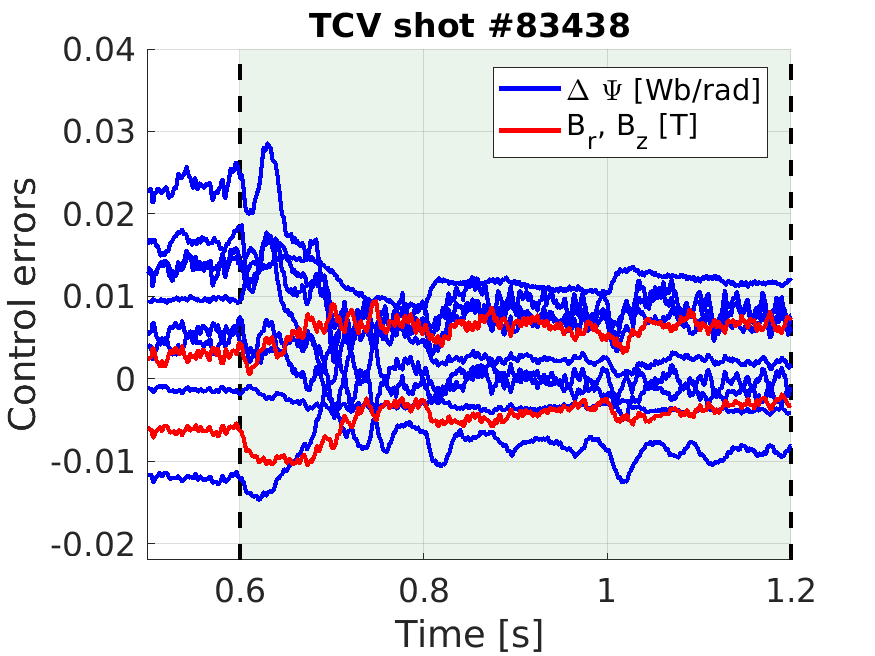}
\caption{Test Case III {(TCV pulse \#83438)} - time traces of the isoflux control errors. {The poloidal flux errors (at contour and strike points) are shown in blue, while the magnetic field errors (at the X-point) are in red.} The beginning and end of the controlled time window are indicated by the dashed black lines.}\label{figure:shapeerrIII}
\end{minipage}
\hfill
\begin{minipage}[c]{0.52\linewidth}
\includegraphics[width=\linewidth]{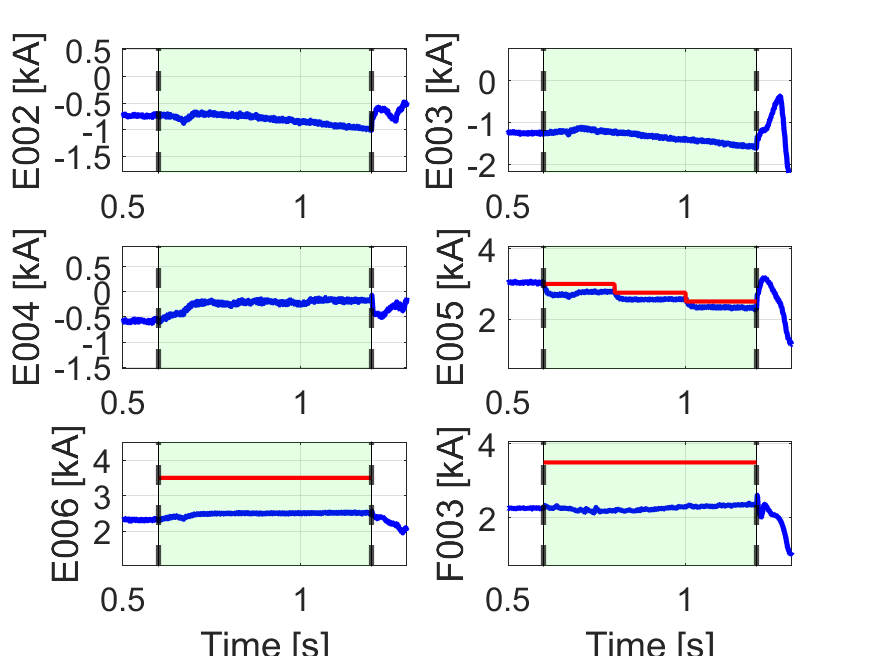}
\caption{Test Case III - time traces of the PF currents (blue) compared with saturation limits (red).}\label{figure:currentIII}
\end{minipage}%
\end{figure}

\begin{figure}
\begin{minipage}[c]{0.48\linewidth}
\includegraphics[width=\linewidth]{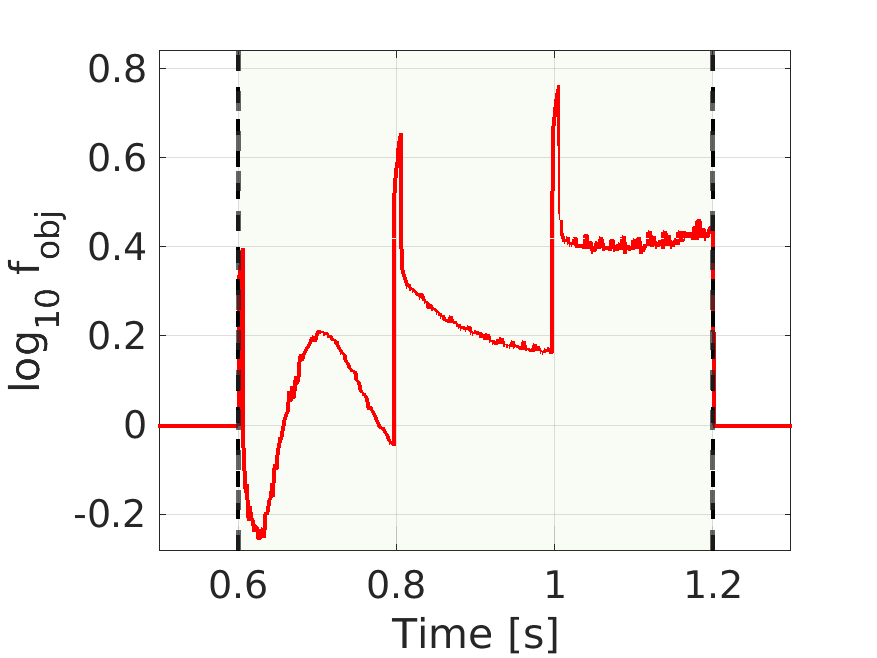}
\caption{Test Case III - time trace of the QP problem objective function.}\label{figure:fobjIII}
\end{minipage}
\hfill
\begin{minipage}[c]{0.48\linewidth}
\includegraphics[width=\linewidth]{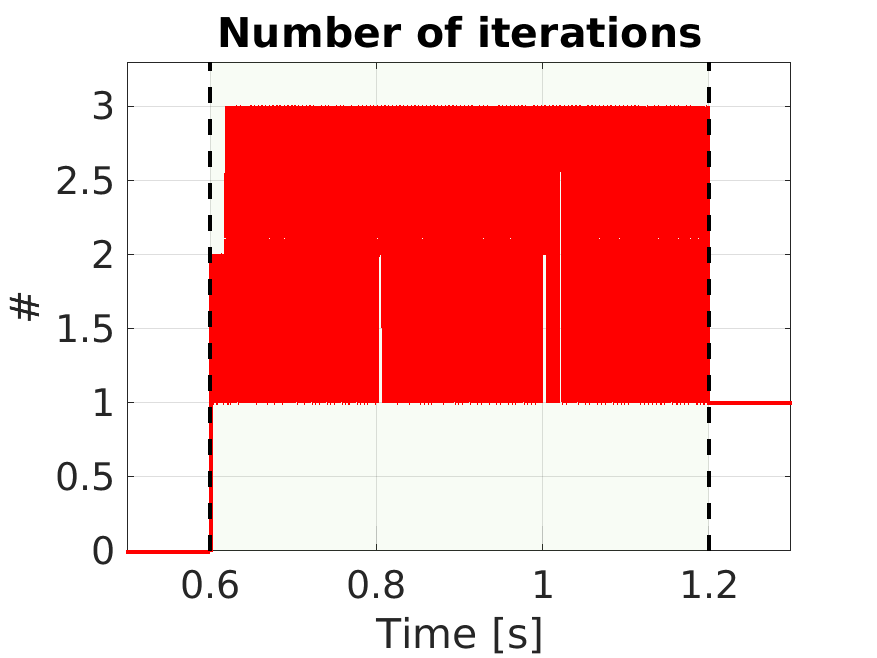}
\caption{Test Case III -iterations performed by the QP solver.}\label{figure:itIII}
\end{minipage}%
\end{figure}

The results for the third test case, TCV shot \#83438, are reported in figs.~\ref{figure:shapeIII}-\ref{figure:fobjIII}. For this case, the saturation on the E5 coils decreases in steps of $250$~A applied every $0.2$~s. 
The snapshots at three different time instants following the CLA activation are shown in Fig.~\ref{figure:shapeIII}, while the time traces of the shape control errors are shown in Fig.~\ref{figure:shapeerrIII}. The shape accuracy degradation is comparable to the one observed in test case II, but in this case the time-varying nature of the saturation limit is clearly reflected by the small jumps observed in the shape control error traces in Fig.~\ref{figure:shapeerrIII} and in the behavior of the QP cost function, reported in~Fig.~\ref{figure:fobjIII}. The relevant coil currents and corresponding saturation limits are reported in~Fig.~\ref{figure:currentIII}, where the variable saturation limit for the E5 current is also shown. The~E5 current returns into the acceptable time-varying band following the activation of the CLA at $t=0.6$. The iterations performed by the QP solver are shown in~Fig.~\ref{figure:itIII}.

{It should be noted that in general time-varying saturations are not expected in tokamaks; in fact, this example was mainly designed as a \vv{stress-test} to showcase the flexibility of the proposed architecture. However, a similar case of practical interest may occur if one wished to introduce constraints on currents that are already beyond the saturation level at the activation time of the CLA system in a gradual fashion, to avoid large transients due to these currents being brought abruptly back into the saturation limits. It can be noticed how, for the considered case, the peak transient shape error in Fig.~\ref{figure:shapeerrIII} is indeed reduced compared to Figs.~\ref{figure:shapeerrI} and~\ref{figure:shapeerrII}.}

\medskip

\section{Conclusive Remarks}\label{section:Conclusions}
This paper describes the implementation of an effective solution to tackle the problem of the~PF coil currents saturation. The methodology proposed for inclusion in the ITER PCS has been tested experimentally on the TCV tokamak. The algorithm proved to be capable of robustly dealing with the saturation of several active circuit currents, { as shown by the} example of section~\ref{subsection:TCII}, and also proved to be capable of handling time-varying saturations, as demonstrated by the example discussed in section~\ref{subsection:TCIII}.

{
It is worth to remark that since the proposed CLA strategy is only related to the expected steady-state behavior, it does not provide any performance guarantee during transients. In fact, the introduction of hard constraints on coil currents could lead to fast variations in the latter ones, which could potentially perturb the plasma configuration. One possibility to mitigate this could be that of introducing the saturation limits gradually, for example as demonstrated by the third proposed example.

However, it should also be noted that the transient performance depends fundamentally on the underlying control architecture, and is not affected by the introduction of the allocator, which only modifies the reference signals to such controller. As it can be seen from the results in section~\ref{section:Exp}, for the considered cases the observed shape degradation during transients remained on an acceptable level. 

As discussed in the introduction to this article, a control technique that is often compared to current allocation algorithms in terms of its capability to handle hard constraints is MPC. Compared to the proposed CLA, MPC has the advantage of offering an integrated solution to deal with both control requirements and constraints, and it naturally provides finer control over transient phases. However, it should be noted that even with MPC, if a hard constraint is activated on a coil whose current is already significantly beyond the saturation threshold, a fast transient is inevitable. 

The proposed CLA, on the other hand, offers two main advantages over MPC. First, since it only deals with steady-state constraints, its implementation is much lighter from a computational point of view. Moreover, especially when a coil current controller is installed on the tokamak, a detailed knowledge of the underlying control architecture or of the plasma dynamics is not necessary.
}

The simplicity of the proposed scheme and the tractable computation cost make this algorithm an effective tool for future reactors such as ITER.

\section*{Acknowledgments}
This work has been supported by the following financial contributions:
\begin{itemize}
    \item Grant Agreement No 101052200 — EUROfusion and training programme 2014-2018 and 2019-2020 under grant agreement No. 633053. 
    \item Swiss State Secretariat for Education, Research and Innovation (SERI). 
    \item TRAINER project (CUP E53D23014670001) funded by EU in NextGenerationEU plan through the Italian “Bando Prin 2022 - D.D. 1409 del 14-09-2022” by MUR
    \item Italian Research Ministry under project PRIN 2022JCZJ33.
\end{itemize}

Opinions expressed are those of the author(s) only and do not necessarily reflect those of the European Union (EU), the European Commission (EC), ITER IO, or SERI. Neither the EU nor the EC nor SERI can be held responsible for them.

\bibliographystyle{spmpsci_unsrt}
\bibliography{CLA.bib}

\end{document}